\documentclass[aps,prb,twocolumn,showpacs,preprintnumbers,amsmath,amssymb,superscriptaddress]{revtex4}%
\usepackage{graphicx}
\usepackage{dcolumn}
\usepackage{bm}
\usepackage{color}

\begin{document}

\preprint{{\em FIRST DRAFT for PRB} (\today)}

\title{Evidence for strong lattice effects as revealed from huge unconventional oxygen isotope effects on the pseudogap temperature in La$_{2-x}$Sr$_{x}$CuO$_{4}$}

\author{M.~Bendele}
\email{markus.bendele@gmail.com} 
\affiliation{Physik-Institut der
Universit\"{a}t Z\"{u}rich, Winterthurerstrasse 190, CH-8057
Z\"{u}rich, Switzerland}
\affiliation{Rome International Center for Materials Science Superstripes (RICMASS), Via dei Sabelli 119A, I-00185 Rome, Italy}

\author{F.~von Rohr}
\affiliation{Physik-Institut der Universit\"{a}t Z\"{u}rich, Winterthurerstrasse 190, CH-8057
Z\"{u}rich, Switzerland}

\author{Z.~Guguchia}
\affiliation{Laboratory for Muon Spin Spectroscopy, Paul Scherrer Institute, CH-5232
Villigen PSI, Switzerland}

\author{E.~Pomjakushina}
\affiliation{Laboratory for Scientific Developments and Novel Materials, Paul Scherrer Institute, CH-5232 Villigen PSI, Switzerland}

\author{K.~Conder}
\affiliation{Laboratory for Scientific Developments and Novel Materials, Paul Scherrer Institute, CH-5232 Villigen PSI, Switzerland}

\author{A.~Bianconi}
\affiliation{Rome International Center for Materials Science Superstripes (RICMASS), Via dei Sabelli 119A, I-00185 Rome, Italy}
\affiliation{Department of Physics of Solid State and Nanosystems, National Research Nuclear University, Kashirskoye shosse 31, 115409, Moscow, Russia}

\author{A.~Simon}
\affiliation{Max-Planck Institute for Solid State Research, Heisenbergstrasse 1, D-70569 Stuttgart, Germany}

\author{A.~Bussmann-Holder}
\affiliation{Max-Planck Institute for Solid State Research, Heisenbergstrasse 1, D-70569 Stuttgart, Germany}

\author{H.~Keller}
\affiliation{Physik-Institut der Universit\"{a}t Z\"{u}rich,
Winterthurerstrasse 190, CH-8057 Z\"{u}rich, Switzerland}

\begin{abstract}
The oxygen isotope ($^{16}$O/$^{18}$O) effect (OIE) on the pseudogap (charge-stripe ordering) temperature $T^{\ast}$ is investigated for the cuprate superconductor La$_{2-x}$Sr$_{x}$CuO$_{4}$ as a function of doping $x$ by means of x-ray absorption near edge structure (XANES)
studies. A strong $x$ dependent and sign reversed OIE on $T^{\ast}$ is observed. The OIE exponent $\alpha_{T^{\ast}}$ systematically decreases from $\alpha_{T^{\ast}} = - 0.6(1.3)$ for $x = 0.15$ to $\alpha_{T^{\ast}} = - 4.4(1.1)$ for $x = 0.06$, corresponding to increasing $T^{\ast}$ and decreasing superconducting transition temperature $T_{c}$.  Both $T^{\ast}(^{16}{\rm O})$ and $T^{\ast}(^{18}{\rm O})$ exhibit a linear
doping dependence with different slopes and critical end points (where $T^{\ast}(^{16}{\rm O})$ and $T^{\ast}(^{18}{\rm O})$ fall to zero) at  $x_{c}(^{16}{\rm O}) = 0.201(4)$ and $x_{c}(^{18}{\rm O}) = 0.182(3)$, indicating a large positive OIE of $x_{c}$ with an exponent of $\alpha_{x_{c}} = 0.84(22)$.
The remarkably large and strongly doping dependent OIE on $T^{\ast}$ signals a substantial involvement of the lattice in the formation of the pseudogap, consistent with a polaronic approach to cuprate superconductivity and the vibronic character of its ground state. 

\end{abstract}
\pacs{74.25-q, 74.72.-h, 74.72.Kf, 74.62.Yb, 61.05.cj}

\maketitle

\section{INTRODUCTION}

Even 30 years after its discovery, the origin of high temperature superconductivity in cuprate superconductors (HTSs) is still under debate \cite{Bednorz86}. A seemingly majority of researchers advocates for a purely electronic mechanism, and consequently ignores the importance of lattice effects \cite{Moriya00}. However, in conventional superconductors the role of the lattice as a pairing glue for Cooper pairs has been identified by the observation of an isotope effect on the superconducting transition temperature $T_{c}$ \cite{Bardeen57}. Interestingly, in cuprate superconductors analogous but apparently unconventional and doping dependent isotope effects have been observed on several characteristic quantities, e.g., the superconducting transition temperature $T_{c}$, the antiferromagnetic transition temperature $T_{N}$, the spin-glass temperature $T_{g}$, the spin-stripe ordering temperature $T_{so}$, and the magnetic penetration depth $\lambda$ \cite{Franck94, Zhao00, Keller05, Schneider05, Keller08, Khasanov08, Weyeneth11, Guguchia14} which all are in support of the vibronic character of its ground state. 
So far, only a few studies of the isotope effect on the pseudogap temperature $T^{\ast}$ have been reported
\cite{Williams98,Raffa98,Lanzara99,Rubio00,Rubio00a,Rubio01,Rubio02,Haefliger06}. 

The pseudogap temperature $T^{\ast}$ plays a central role in the phase diagram of HTS's and is considered to be a key ingredient for the understanding of the physics of cuprates \cite{Lee06}. From an experimental point of view $T^{\ast}$ itself is a rather ill-defined temperature scale with varying values depending on the experimental tool and the involved time scale \cite{Williams98,Raffa98, Lanzara99, Rubio00}. The nature of  $T^{\ast}$ 
in cuprates is debated and attributed to controversial origins \cite{Lee14}. These are discussed as precursors to the paired state, spin density formation, charge density formation, onset of stripe formation and other exotic scenarios \cite{Lee14}. Here we define the pseudogap $T^{\ast}$ as the temperature where lattice effects are apparent in terms of deviations of the local structure from the average one (this temperature is also denoted as 
charge-stripe ordering temperature in the literature \cite{Lanzara99}).
The relevance of lattice/polaron effects for the appearance of the pseudogap phase can be tested by oxygen isotope ($^{16}$O/$^{18}$O) effect (OIE) studies. So far, only a limited number of such experiments have been performed \cite{Williams98,Raffa98, Lanzara99,Rubio00, Rubio00a, Rubio01,Rubio02,Haefliger06}. First $^{89}$Y nuclear magnetic resonance (NMR) measurements on slightly underdoped YBa$_{2}$Cu$_{4}$O$_{8}$ show no appreciable OIE on $T^{\ast}$ \cite{Williams98}, whereas $^{63}$Cu nuclear quadrupole resonance (NQR) spin-lattice relaxation experiments on the same  system revealed a small but positive OIE on $T^{\ast}$ \cite{Raffa98}. 
However, a large and sign reversed OIE has been reported from Cu K-edge x-ray absorption near edge structure (XANES) studies in underdoped La$_{1.94}$Sr$_{0.06}$CuO$_{4}$ \cite{Lanzara99}, and from neutron crystal field spectroscopy (NCFS) studies in slightly underdoped HoBa$_{2}$Cu$_{4}$O$_{8}$ \cite{Rubio00}, optimum doped La$_{1.81}$Ho$_{0.04}$Sr$_{0.15}$CuO$_{4}$ \cite{Rubio02}, and doped La$_{1.96-x}$Ho$_{0.04}$Sr$_{x}$CuO$_{4}$ $(0.11 \leq x \leq 0.25)$ \cite{Haefliger06}.
 Moreover, NCFS studies of the $^{63}$Cu/$^{65}$Cu isotope effect evidenced a large negative isotope shift of  $T^{\ast}$ in HoBa$_{2}$Cu$_{4}$O$_{8}$ \cite{Rubio01}, whereas no copper isotope shift of  $T^{\ast}$ was observed for optimum doped La$_{1.81}$Ho$_{0.04}$Sr$_{0.15}$CuO$_{4}$ \cite{Rubio02}.

Here we present a systematic study of the OIE on $T^{\ast}$ over a broad doping range of La$_{2-x}$Sr$_{x}$CuO$_{4}$ ($0.06 \leq x \leq 0.15$) by means of Cu K-edge XANES studies \cite{Li91}. 

\section{EXPERIMENTAL DETAILS}

Four different La$_{2-x}$Sr$_{x}$CuO$_{4}$ samples with concentrations $x =0.06, 0.09, 0.12, 0.15$ were prepared by a conventional solid state reaction using dried La$_{2}$O$_{3}$ (99.99\%), SrCO$_{3}$ (99.999\%), and CuO (99.999\%). The powders were mixed, ground thoroughly, pressed into pellets and fired in air at 1000 $^{\circ}$C for $\sim$ 100~h with three intermediate grindings. The phase purity of the samples was checked with a conventional x-ray diffractometer. The final $^{18}$O and $^{16}$O samples were obtained in parallel via oxygen isotope exchange in closed quartz glass tubes under controlled gas pressure slightly above 1 bar at 800 $^{\circ}$C during 72 hours with subsequent 
cooling. The oxygen-isotope enrichment was determined from the weight changes and was about 90\% of $^{18}$O for all samples. 

%
\begin{figure}[t]
\centering
\includegraphics[width=0.75\linewidth]{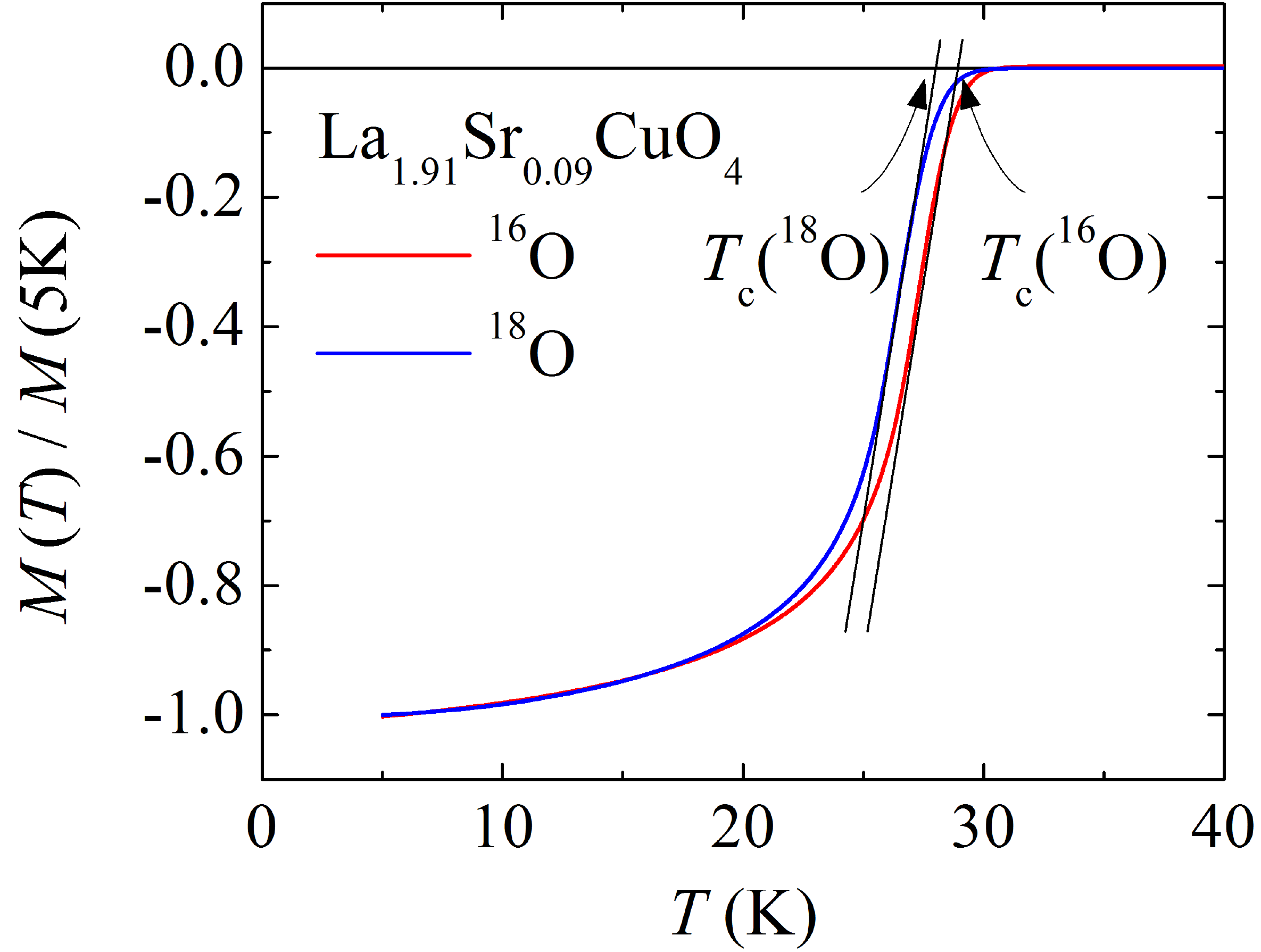}
\caption{(Color online) Temperature dependence of the zero-field cooled normalized magnetization $M(T)/M(5K)$ of the $^{16}$O/$^{18}$O exchanged samples of La$_{1.91}$Sr$_{0.09}$CuO$_{4}$ in $\mu_{0}H = 0.5~{\rm mT}$. The superconducting transition temperature was determined by the intersection of the linearly extrapolated magnetic moment and the zero line.} 
\label{Fig1_mag_OIE_Tc}
\end{figure}
%

The samples were characterized by measuring their superconducting properties using a SQUID magnetometer ({\em Quantum Design MPMS-XL}). 
The temperature dependencies of the zero field cooled normalized magnetization $M(T)/M(5K)$ of the $^{16}$O/$^{18}$O exchanged samples of La$_{1.91}$Sr$_{0.09}$CuO$_{4}$ for $\mu_{0}H = 0.5~{\rm mT}$ are shown in 
Fig.~\ref{Fig1_mag_OIE_Tc} with $T_{c}$ being identified from the zero of the linearly extrapolated magnetic moment. From the values of 
$T_{c}(^{16}{\rm O}) $ and $T_{c}(^{18}{\rm O}$) the OIE exponent  $\alpha_{T_{c}}$ is derived according to 
$\alpha_{T_{c}} =  - {\rm d}\ln T_{c}/{\rm d}\ln M_{O}$, where $M_{O}$ is the mass of the oxygen isotope. 
The results of the OIE on $T_{c}$ for La$_{2-x}$Sr$_{x}$CuO$_{4}$ ($0.06 \leq x \leq 0.15$) will be discussed in more detail below, in connection with the OIE on $T^{\ast}$ observed in this work. 

In order to ascertain $T^{\ast}$, Cu K-edge XANES measurements were performed on powder samples at the Diamond Light Source on the B18 beamline \cite{Dent13}. The storage ring was operating in a 10-minute top-up mode for a ring current of 300 mA and an energy of 3 GeV. The radiation was monochromated with a Si (311) double crystal monochromator calibrated using the K-edge of a copper foil, taking the first inflection point as 8979.0 eV. Harmonic rejection was achieved through the use of two platinum-coated mirrors operating at an incidence angle of 7.5~mrad. 
 The incident ($I_{0}$), the transmitted ($I_{t}$), and the reference ($I_{\rm ref}$) beam intensities were measured using two 30 cm ion chambers and a helium/argon gas mixture (total ion chamber pressure of 1.1 bar) to absorb 12\%, 60\%, and 60\% of the beam in $I_{0}$, $I_{t}$ and $I_{\rm ref}$, respectively \cite{Dent13}.The samples were positioned between $I_{0}$ and $I_{t}$ and the Cu foil reference is between $I_{t}$ and $I_{\rm ref}$.

\begin{figure}[h]
\centering
\includegraphics[width=0.65\linewidth]{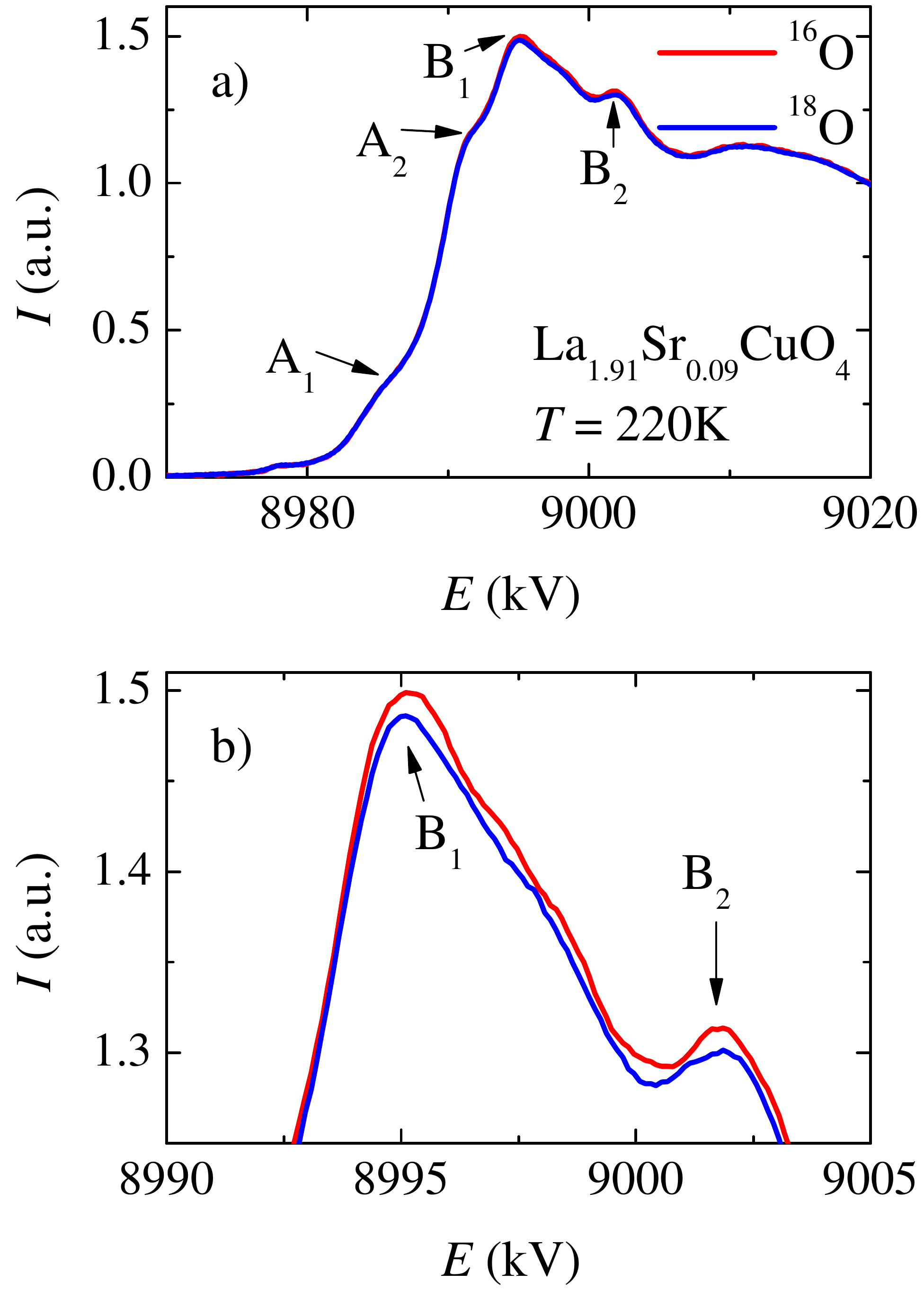}
\caption{(Color online) a) Normalized Cu K-edge XANES spectra showing the energy evolution of the normalized intensity $I$ for both isotopes $^{16}$O and $^{18}$O of La$_{1.91}$Sr$_{0.09}$CuO$_{4}$ at $T = 220~{\rm K}$. b) Extended region of the XANES spectrum depicted in panel (a) showing a slight change in the shape due to oxygen isotope substitution ($^{18}$O/$^{18}$O).} 
\label{Fig2_XANES_spectra}
\end{figure}

Five absorption spectra were collected at the same temperature in an {\em Oxford Instruments OptistatPT} cryostat and their mean value taken in order to increase the signal to noise ratio. Figure \ref{Fig2_XANES_spectra} shows the normalized Cu K-edge XANES spectra for the $^{16}$O and $^{18}$O La$_{1.91}$Sr$_{0.09}$CuO$_{4}$ samples at $T = 220~{\rm K}$ as a representative for the entire series. The spectra show the typical absorption features of the cuprates with a square planar geometry where the main features are denoted as $A_{1}$, $A_{2}$, $B_{1}$, and $B_{2}$ with peak intesities $a_{1}$, $a_{2}$, $b_{1}$, and $b_{2}$. The peak $A_{1}$ is essentially caused by the multiple scatterings of the ejected photoelectrons from the La/Sr atoms, while peak $A_{2}$ is due to the apical oxygen. The features above the threshold from 8 eV to 40 eV have been identified as multiple scattering resonances within a 
0.5~nm cluster of neighbor atoms surrounding the absorbing atom \cite{Li91}, i.e. $B_{1}$ probes the scattering from the oxygen atoms in the square CuO$_{2}$ planes. 
The decrease of the $b_1$ and $b_2$ resonance intensities together with the increase of both $A_1$ and $A_2$ multiple scattering photoelectron resonance intensities $a_1$ and $a_2$, indicate deviations from the CuO$_2$ square plane geometry for anisotropic atomic displacements in the CuO$_4$ plaquette \cite{Li91}, which has been used to probe the onset of periodic local lattice distortions associated with polaron formation.

%
%
%
\begin{figure}[h]
\centering
\includegraphics[width=1.0\linewidth]{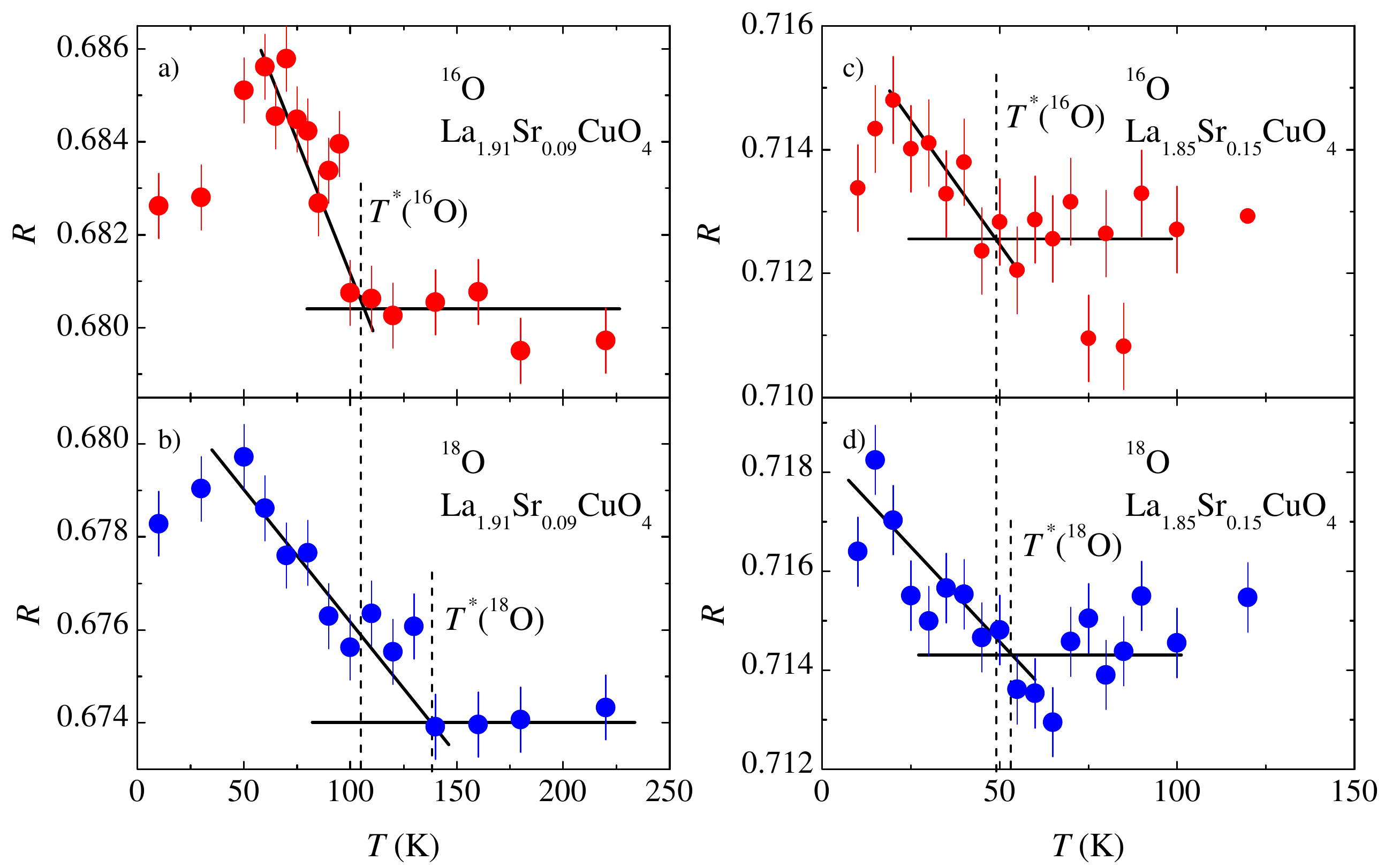}
\caption{(Color online) The temperature dependence of the XANES peak intensity ratio $R$ of oxygen-isotope substituted La$_{2-x}$Sr$_{x}$CuO$_{4}$: $x = 0.09$: (a) $^{16}$O and (b) $^{18}$O; $x = 0.15$: (c) $^{16}$O and (d) $^{18}$O. The solid straight lines are a guide to the eye. The corresponding charge stripe ordering temperatures $T^{\ast}$ are marked by the vertical dashed lines. The huge OIE on $T^{\ast}$ is evident.} 
\label{Fig3_OIE _shift_T*}
\end{figure}
%
The effect of the oxygen-isotope substitution on the XANES spectra on the distribution of local Cu-site conformations is hardly seen directly in the raw XANES spectra shown in Fig. \ref{Fig2_XANES_spectra}, but can be extracted from the slightly different XANES peak intensities $a_1$ and $b_1$ of  the 
$^{16}$O and  $^{18}$O  samples \cite{Lanzara99}. Deviations from the square plane geometry of the CuO$_{4}$  coordination plaquette can be quantified by the XANES peak ratio \cite{Lanzara99}
\begin{equation}
R = (b_{1} - a_{1})/(b_{1} + a_{1}) \, .
\end{equation}
In Fig.~\ref{Fig3_OIE _shift_T*} the XANES intensity ratio $R$ is plotted for $^{16}$O and $^{18}$O samples
of La$_{2-x}$Sr$_{x}$CuO$_{4}$ ($x = 0.9$ and $x = 0.15$) as a function of temperature.
From the temperature dependence of $R$, the pseudogap temperature $T^{\ast}$ is determined by the interception of the horizontal line above $T^{\ast}$ and to the linear slope below $T^{\ast}$ as shown in Fig.~\ref{Fig3_OIE _shift_T*}. For all doping levels $x$, a clear increase in $R$ below $T^{\ast}$ is observed which is typical for a charge density wave and consistent with recent resonant X-ray scattering studies, where charge ordering and the pseudogap formation were intensively studied \cite{Comin16}.
Below $T_{c}$ the quantity $R$ decreases sharply caused by the decrease of charge carriers due to the formation of the superconducting condensate.

%
\begin{figure}[h!]
\centering
\includegraphics[width=0.7\linewidth]{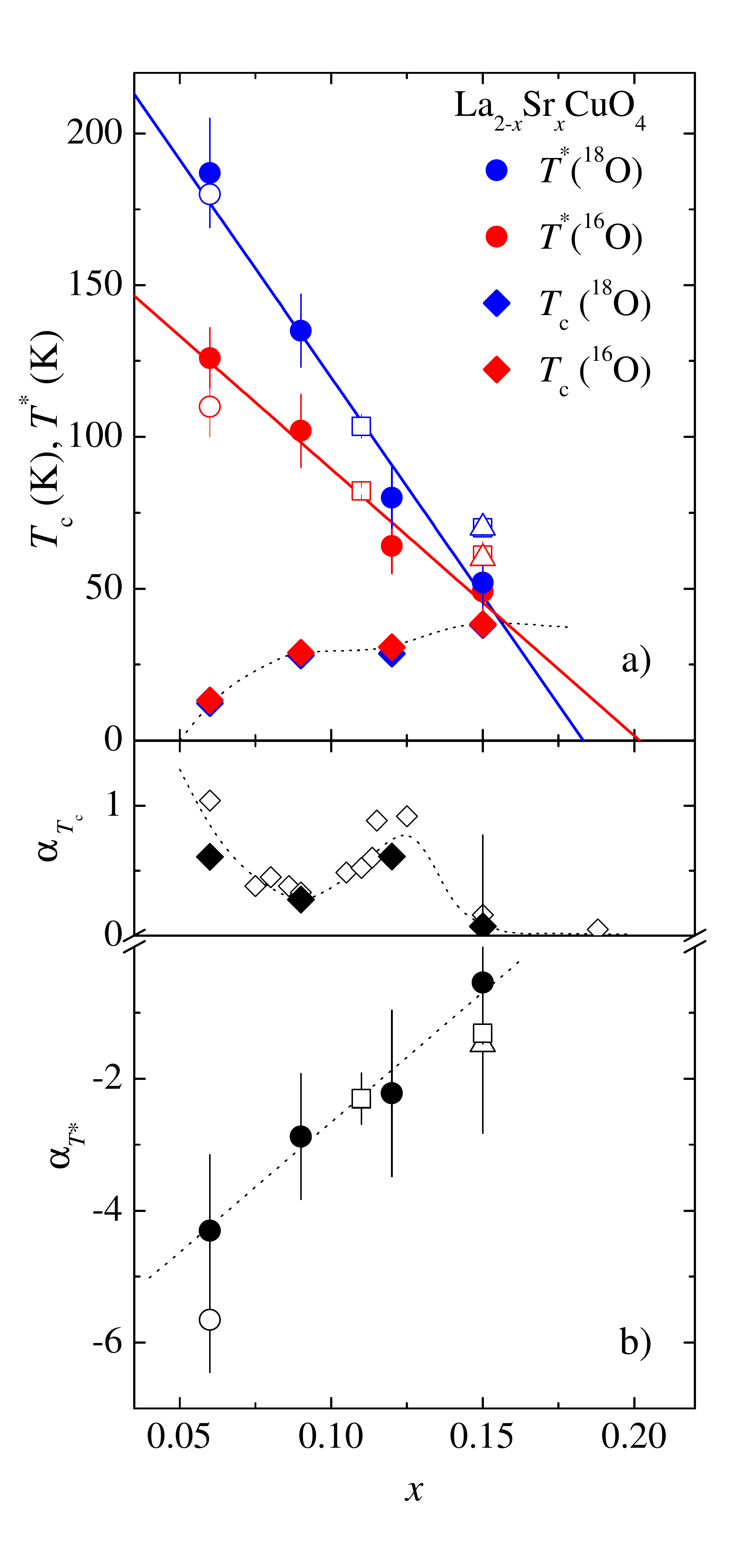}
\caption{(Color online) a) The superconducting transition temperature $T_{c}$ and the pseudogap temperature $T^{\ast}$ of La$_{2-x}$Sr$_{x}$CuO$_{4}$ as a function of doping $x$ for $^{16}$O (red symbols) and $^{18}$O (blue symbols). With increasing doping $x$ a continuous decrease of  $T^{\ast}$ is observed for both isotopes. The solid lines are obtained from Eq.~(\ref{T*_x}). 
The dashed line is a guide to the eye. For comparison, the values for $T^{\ast}(^{16}{\rm O})$ and $T^{\ast}(^{18}{\rm O})$ obtained from previous XANES experiments for La$_{1.94}$Sr$_{0.06}$CuO$_{4}$ (open circles) \cite{Lanzara99} and from previous NCFS experiments for La$_{1.96-x}$Ho$_{0.04}$Sr$_{x}$CuO$_{4}$ ($x = 0.15$  (open triangles) \cite{Rubio02}; $x = 0.11, 0.15$ (open squares) \cite{Haefliger06}) are also displayed. 
b) Doping dependence of the isotope effect exponent
$\alpha_{T_{c}}$ (closed diamonds) and  $\alpha_{T^{\ast}}$ (closed circles) for La$_{2-x}$Sr$_{x}$CuO$_{4}$ as determined in this work, together with published values for $\alpha_{T_{c}}$ (open symbols) \cite{Crawford90} and values for $\alpha_{T^{\ast}}$ extracted from the previous XANES and NCFS results shown in panel (a): XANES (open circle) \cite{Lanzara99}; NCFS (open triangle) \cite{Rubio02}, and NCFS (open squares) \cite{Haefliger06}. The dashed lines are a guide to the eye.}
\label{Fig4_OIE_T*}
\end{figure}

\section{RESULTS}

The values of the pseudogap temperature $T^{\ast}$ as well as the corresponding values of $T_{c}$ as measured for all $^{16}$O and $^{18}$O samples of La$_{2-x}$Sr$_{x}$CuO$_{4}$ ($0.06 \leq x \leq 0.15$) are summarized in Fig.~\ref{Fig4_OIE_T*}a and Table~\ref{table_T*}. 
$T_{c}(x)$ shows the well-known dome-like behavior with a small, but clearly visible OIE on $T_{c}$.
$T^{\ast}$ decreases with increasing $x$ as is well known from the literature. The two isotopes show the same trends with $x$, however, 
$T^{\ast}(^{18}{\rm O})$  being pronouncedly raised as compared to $T^{\ast}(^{16}{\rm O})$
with the largest OIE shift of $\Delta T^{\ast} \simeq 61~{\rm K}$ for $ x = 0.06$. With increasing $x$, the OIE on $T^{\ast}$ decreases systematically. For comparison, the values for $T^{\ast}(^{16}{\rm O})$ and $T^{\ast}(^{18}{\rm O})$ obtained from previous XANES experiments for La$_{1.94}$Sr$_{0.06}$CuO$_{4}$  \cite{Lanzara99} and from previous NCFS experiments for La$_{1.96-x}$Ho$_{0.04}$Sr$_{x}$CuO$_{4}$ ($x = 0.15 $ \cite{Rubio02}; $x = 0.11, 0.15$ \cite{Haefliger06}) are also displayed in Fig.~\ref{Fig4_OIE_T*}a. These previous values for $T^{\ast}(^{16}{\rm O})$ and $T^{\ast}(^{18}{\rm O})$ are within experimental error in good agreement with the present results, demonstrating that the two different experimental techniques XANES and NCFS yield consistent values for $T^{\ast}$. 
Note that for overdoped La$_{1.96-x}$Ho$_{0.04}$Sr$_{x}$CuO$_{4}$ ($x = 0.20$ and $x = 0.25$) NCFS experiments \cite{Haefliger06} revealed in contrast to other experiments an unexpected upward shift of $T^{\ast}$ with an appreciable negative OIE on $T^{\ast}$ which, however, is not discussed here.

The  corresponding OIE exponent $\alpha_{T^{\ast}}$ is evaluated analogously to the one on $T_{c}$. It is negative and unusually large with $\alpha_{T^{\ast}}=  - 4.4(1.1)$ for the lowest doping $x = 0.06$. The magnitude of $\alpha_{T^{\ast}}$ is decreasing with increasing $x$, reaching a still substantial value of $\alpha_{T^{\ast}} = - 0.6(1.3)$ at optimal doping $x = 0.15$ (see Table~\ref{table_T*}). The doping dependencies
of the OIE exponents $\alpha_{T^{\ast}}$ and $\alpha_{T_{c}}$ of La$_{2-x}$Sr$_{x}$CuO$_{4}$ 
($0.06 \leq x \leq 0.15$) are displayed in Fig.~\ref{Fig4_OIE_T*}b (full symbols), together with previous results (open symbols: $\alpha_{T^{\ast}}$ \cite{Lanzara99,Rubio02,Haefliger06}, $\alpha_{T_{c}}$ \cite{Crawford90}) which are confirmed here.

\section{DISCUSSION}

It is evident from Fig.~\ref{Fig4_OIE_T*}a that in the underdoped regime ($0.5 \leq x \leq 0.15$) the pseudogap temperatures $T^{\ast}(^{16}{\rm O})$ and $T^{\ast}(^{18}{\rm O})$ exhibit almost linear doping dependences which can be described by the expression:
\begin{equation}
T^{\ast}(^{\gamma}{\rm O}) = \sigma (^{\gamma}{\rm O}) [x - x_{c}(^{\gamma}{\rm O})], \,\, \gamma = 16, 18
\label{T*_x}
\end{equation}
Here $\sigma (^{\gamma}{\rm O})$ and  $x_{c}(^{\gamma}{\rm O})$ are constants. The present $T^{\ast}$ data (Fig.~\ref{Fig4_OIE_T*}a and Table~\ref{table_T*}) follow Eq.~(\ref{T*_x}), yielding 
$\sigma (^{16}{\rm O}) = - 880(100) {\rm K}$, $x_{c}(^{16}{\rm O}) = 0.201(4)$ and $\sigma (^{18}{\rm O}) = - 1440(170) {\rm K}$, $x_{c}(^{18}{\rm O}) = 0.182(3)$. It is interesting to note that both quantities $\sigma$ and $x_{c}$ exhibit a substantial OIE. The physical meaning and the location of the $T^{\ast}$ line in the phase diagram (temperature versus hole doping) of cuprates have extensively been discussed in the literature and are still heavily debated (see, e.g. \cite{Varma97,Sachdev00,Tallon01,Norman05,Huefner08,Li08,Shekter13,Keimer15}). A linear doping dependence of $T^{\ast}$ has previously been reported for various cuprate systems, and the extrapolated generic doping concentration $x_{c} \simeq 0.19 - 0.20$ has been discussed in terms of a possible quantum critical point underneath the superconducting dome (see, e.g. \cite{Varma97, Sachdev00,Tallon01,Li08,Shekter13}). Note that $x_{c}$ decreases from 0.201(4) to 0.182(3) upon replacing $^{16}$O by $^{18}$O, implying a large positive OIE of about 10\%, corresponding to an OIE exponent of $\alpha_{x_{c}} = 0.84(22)$ (see Fig.~\ref{Fig4_OIE_T*}a). At present, a theoretical interpretation of this remarkable OIE ist not available.

%
%
%
%
\begin{table}[h]
\caption{Summary of the values of  the pseudogap temperatures $T^{\ast} (^{16}{\rm O})$, $T^{\ast} (^{18}{\rm O})$, and the OIE exponent $\alpha_{T^{\ast}}$ for La$_{2-x}$Sr$_{x}$CuO$_{4}$ ($0.06 \leq x \leq 0.15$).}
\vspace{0.3cm}
\begin{tabular}{lccl}
%
\hline
\hline  
Doping $x$        &  $T^{\ast} (^{16}{\rm O})$ [K] & $T^{\ast} (^{18}{\rm O})$ [K] & $\alpha_{T^{\ast}}$ \\ \hline
0.06 & 126(10) & 187(18)  &  - 4.3(1.1)\\ 
0.09 & 102(12) & 135(12)  & - 2.9(9) \\ 
0.12 & 64(9) & 80(10) & - 2.2(1.3) \\ 
0.15 & 49(8) & 52(8) & - 0.6(1.3) \\ 
\hline
\hline 
\end{tabular}
\label{table_T*}
\end{table}
%
%

The OIE on $T_{c}$ has been consistently explained earlier using a polaronic approach 
\cite{Bussmann05,Bussmann07,Keller08,Weyeneth11,Muller07,Muller14}, and within multi-gap multi-condensate superconductivity at the BEC-BCS crossover near a Lifshitz transition in multi-component electronic systems \cite{Bianconi97,Perali12}. It is worth mentioning that in an early work Kresin and Wolf \cite{Kresin94} already proposed to explain the OIE on $T_{c}$ and on the penetration depth, considering a strong $c$-axis electron-lattice coupling.  More recently this model was extended by Weyeneth and M\"{u}ller \cite{Weyeneth11, Muller14} to include a polaronic coupling also to the planes. They succeeded in quantitatively reproducing the experimental data thus demonstrating the importance of polaron formation for superconductivity in the cuprates. 

The basic ingredients of polaron formation are that strong electron-lattice interactions lead to polaron formation, namely quasiparticles, which renormalize the electronic as well as the lattice degrees of freedom. The squeezing of the electronic hopping integrals has been identified as the origin of the OIE on $T_{c}$ \cite{Keller08}.
Here we argue that the OIE on $T^{\ast}$ is caused by polaron formation related to an incipient finite momentum lattice instability which manifests itself as a divergence in Cu-O mean-square displacement at  $T^{\ast}$ as has been observed in \cite{Bianconi96,Oyanagi09}. 
%

The primary semi-classical response of the lattice to polaron formation is a rigid
harmonic oscillator shift, {\em i.e.}, the phonon creation and annihilation operators $b^{+}$ and $b$ transform to
\begin{equation}
\tilde{b}_{q} = b_{q} + \sum_{q} \gamma_{i}(q) \, c^{+}_{i}c_{i}\,\,\,, 
\tilde{b}^{+}_{q} = b^{+}_{q} + \sum_{q} \gamma_{i}(q)\,  c^{+}_{i}c_{i}\, ,
\end{equation}
with $c^{+}_{i}$ and $c_{i}$ being site $i$ dependent electron creation and annihilation operators and $\gamma_{i}(q)$ the momentum $q$
dependent coupling constants. As a consequence of this coupling the lattice mode
frequencies adopt a temperature dependence \cite{Bussmann05, Bussmann07, Keller10}:
\begin{equation}
\tilde{\omega}_{q,j}^2 =  \tilde{\omega}_{0q,j}^2 - \frac{\gamma_{q,j}^2}{N(E_{F})} \sum_{k} \frac{1}{\varepsilon (k)}\tanh \frac{\varepsilon(k)}{k_{B}T} \,\,,
\end{equation}
where $N(E_{F})$ is the density of states at the Fermi level, $\varepsilon(k)$ the Fourier transform
of the site representation of the electronic bands, and $\tilde{\omega}_{0q,j}$ the bare branch $j$ and
momentum $q$ dependent lattice mode frequency. The electronic dispersions are
explicitly given by a LDA derived form, namely:
\begin{eqnarray}
\varepsilon(k)  & =  & - 2 t_1[\cos(k_{x}a) + \cos(k_{y}b)]  + 4 t_2 \cos(k_{x}a) \cos(k_{y}b) \nonumber \\
& +  & 2 t_3[\cos(2k_{x}a) + \cos(2k_{y}b)] \nonumber \\
& \mp & t_4 [\cos(k_{x}a) - \cos(k_{y}b)]^2/4 - \mu \,\, ,
\end{eqnarray}
where $t_1$, $t_2$, $t_3$, and $t_4$ are first, second, third, and interlayer hopping integrals, respectively,
and $\mu$ is the chemical potential which controls the band filling and is set equal to zero. Here, $a$ and 
$b$, are the lattice constants in the CuO$_{2}$ planes. It is
important to note that also the electronic degrees of freedom are renormalized which
leads to an exponential band narrowing for all hopping elements according to the
replacement  $t \rightarrow t \exp[- \gamma^2 \coth(\hbar \omega/ 2k_BT)]$ and a rigid level shift.

From Eq. (4) {\em finite momentum mode softening} can occur if the normal mode
unrenormalized frequency $\tilde{\omega}_{0q,j}$ is reduced by the electronic energy, corresponding to
$\tilde{\omega}_{q,j}^2 \rightarrow 0$. Obviously, this situation is controlled by the dependence of the coupling
strength which we use as a variable to calculate the mode freezing temperature where $\tilde{\omega}_{q,j}^2 = 0$.
Since the electronic energies due to their coupling to the lattice degrees of freedom and the unrenormalized 
mode energies are both isotope dependent, the freezing temperature is isotope dependent as well. 
This dependence is huge and sign reversed \cite{Keller10} as compared to the isotope effect on the harmonic lattice mode frequencies and
mirrors the isotope effect on the pseudogap onset temperature $T^{\ast}$ which is identified as
the freezing temperature. $T^{\ast}$ signals the onset of a dynamically modulated, patterned
structure (composed of local coherent polarons), where the superstructure modulation is
defined by the $q$-value where $\tilde{\omega}_{q,j}^2 = 0$ and $T^{\ast}$ is defined through the implicit relation:
\begin{equation}
\tilde{\omega}_{0q,j}^2 = \frac{\gamma_{q,j}^2}{N(E_{F})} \sum_{k} \frac{1}{\varepsilon (k)}\tanh \frac{\varepsilon(k)}{k_{B}T^{\ast}} \,\,.
\end{equation}

%
The Cu-O mean square displacement, which is momentum $q$ and branch $j$ dependent, can be expressed like:
\begin{equation}
\sigma^{2}(T) = \hbar /(M \tilde{\omega}_{q,j}) \coth (\hbar \tilde{\omega}_{q,j} / 2k_{B}T) \, ,
\end{equation}
where $M$ is the oxygen ion mass \cite{Keller10}. As has been shown in \cite{Keller10} the OIE on $T_{c}$ is caused by the renormalization of the band energies where specifically the second nearest neighbor hopping integral yields the correct OIE. This suggests a substantial involvement of a $Q_{2}$-type Jahn-Teller mode in the formation of the pseudogap phase which has already been deduced from the early data for the OIE on $T^{\ast}$ \cite{Rubio00,Rubio01,Rubio02}.
This is further supported by the observation of a $^{63}$Cu/$^{65}$Cu isotope effect which is present only in HoBa$_{2}$Cu$_{4}$O$_{8}$ \cite{Rubio01}, but absent in La$_{1.81}$Ho$_{0.04}$Sr$_{0.15}$CuO$_{4}$ \cite{Rubio02}, since in the former an umbrella type mode including Cu ion displacement is active which is absent in La$_{2-x}$Sr$_{x}$CuO$_{4}$.
Alternatively, the OIE on $T^{\ast}$ (but not on $T_{c}$) has been explained within a purely electronic model where lattice effects were built in at
specific momentum values \cite{Andergassen01}. 

In the above we have not addressed the question why the $^{63}$Cu NQR relaxation study \cite{Raffa98}, opposite to XANES \cite{Lanzara99} and NCFS \cite{Rubio00,Rubio00a,Rubio01,Rubio02, Haefliger06}, detects only a small OIE on  $T^{\ast}$ with reversed sign as compared to the present results. 
First, the $T_{1}$ spin-lattice relaxation measurements performed by NQR probe both magnetic and charge fluctuations. However, magnetic fluctuations dominate in the temperature region investigated. Second, the probing time scales of both techniques are quite different, namely NQR is slow and works in the kHz to MHz range, whereas XANES is fast (10$^{-15}{\rm s}$) and insensitive to magnetism. 
This means that NQR and XANES do not probe the same physical quantities and thus the results obtained for the pseudogap are not directly comparable to each other.

\section{CONCLUSION}

In conclusion, we have performed a systematic investigation of the oxygen isotope effect on the pseudogap temperature
$T^{\ast}$ by using XANES and thereby demonstrated its doping dependent evolution. The isotope effect is
always sign reversed and large and increases with decreasing doping. The correct trends of the
isotope effect are obtained within a polaronic model where the renormalization of lattice and
electronic degrees of freedom is taken into account to calculate the isotope effect on $T^{\ast}$. The present results support the original concept leading to the discovery of high-temperature superconductivity in the cuprates \cite{Bednorz86} that a $Q_{2}$-type
Jahn-Teller mode is at work in the formation of the pseudogap phase as has been concluded previously in explaining the OIE on $T_{c}$
\cite{Muller07,Muller14}. In addition, the data substantiate the notion of the vibronic character of the ground state.

\section*{ACKNOWLEDGEMENTS}

It is a pleasure to acknowledge the year long support from K.A. M\"{u}ller who also suggested to perform this work already several years ago. In addition, numerous encouraging discussions with him are gratefully recognized.  We also thank B. Joseph, M. Mali, C. Marini, and J. Roos for helpful discussions. Financial support by the Swiss National Science Foundation is gratefully acknowledged. We thank Diamond Light Source for access to beamline B18 (proposal number: SP9960) that contributed to the results presented here.
%
%
%

%
\end{document}